\documentclass[prb,showpacs,twocolumn,amsmath,amssymb]{revtex4}
\usepackage{epsfig}
\usepackage{color}
\usepackage{epsf}
\usepackage{amssymb, amsmath}
\usepackage{graphicx}
\usepackage{dcolumn}
\usepackage{bm}

\begin{document}

\title{Chern band insulators in magnetic field
}

\author{
Miguel A. N. Ara\'ujo$^{1,2}$, Eduardo V. Castro$^{1}$}

\affiliation{$^1$ CFIF, Instituto Superior
T\'ecnico, Universidade de Lisboa, Av. Rovisco Pais, 1049-001 Lisboa, Portugal}
\affiliation{$^2$ Departamento de F\'{\i}sica,  Universidade de \'Evora, P-7000-671, \'Evora, Portugal}

\begin{abstract}
The effect of  a magnetic field on a two-dimensional 
Chern band insulator is discussed. It is shown that, 
unlike the trivial insulator, an anomalous Hall insulator  
with Chern  number $C$ becomes a metal when a magnetic field is applied 
at constant particle density, for any $C>0$. For a time reversal invariant
topological insulator 
with a spin Chern resolved number, $C_\uparrow= - C_\downarrow=C$, 
the  magnetic field induces a  spin polarized spin Hall insulator.
We consider also the effect of a superlattice potential and extend 
previous results for the quantization of the Hall conductance of filled 
Hofstadter bands to this problem.   
\end{abstract}


\maketitle

\section{Introduction}

\begin{figure}
\begin{centering}
\includegraphics[width=6cm]{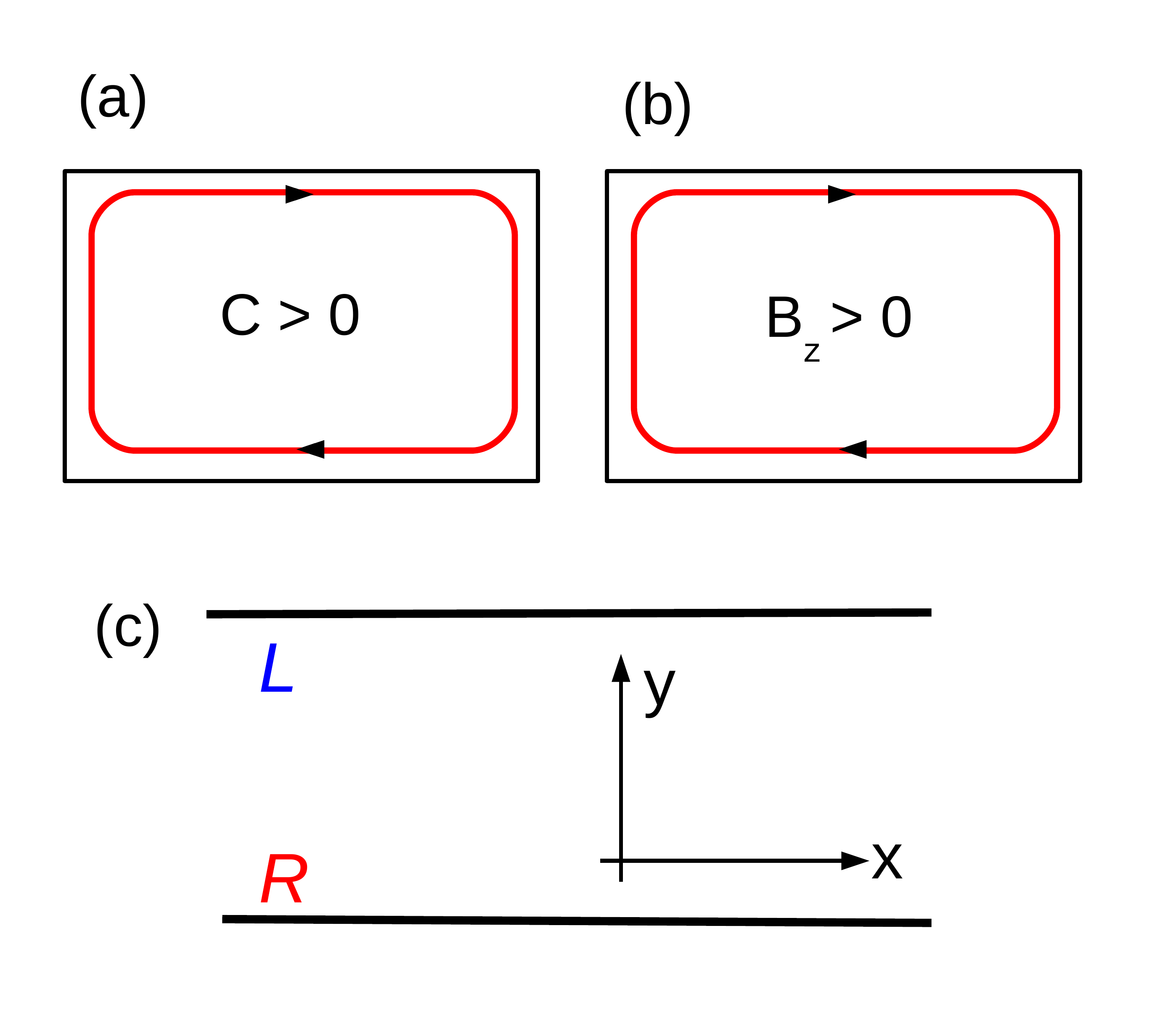}
\par\end{centering}
\caption{\label{chiralities}(color online). The chirality of the edge states
for: (a) a 2D topological system; (b)
a 2D system under an applied perpendicular magnetic field;
(c) the geometry considered in figure \ref{edgestates}.
The chiralities in (a) and (b) are reversed for opposite signs of $C$ or $B_z$.}
\end{figure}

 Two-dimensional Bloch bands with non-trivial topology have
recently become a topic of intense research activity\cite{hasankane,QiZhang}.
For spinless fermions, the most important topological index is the Chern
number, $C$,
of the  filled band. Non-zero $C$ implies the existence  of chiral states
along the system's boundary,
the number of which is given by the bulk-boundary correspondence,
$N_R-N_L=\delta C$, where $N_{R(L)}$ counts the number of right- (left-) moving states along
the boundary between two regions where the Chern number differs by $\delta C$.
If  the system is in contact with a trivial insulator (the vacuum, for instance),
the edge states' chirality depends on the sign of $C$, as shown in figure~\ref{chiralities}(a).
Let us consider now that a  magnetic field is applied perpendicular to a two-dimensional system.
We do not consider here the Zeeman coupling to the spin, only the effect of the
minimal coupling to orbital degrees of freedom.
The  magnetic field's vector potential
 introduces a chirality which manifests itself in the Hall effect (figure~\ref{chiralities}(b)).
The effect of a magnetic field on a two-dimensional Bloch band has long been established:
it splits the original band into subbands, the so-called Hofstadter butterfly spectrum, as
shown explicitly for the square\cite{hofstadter,macdonald}, hexagonal\cite{claro} and
honeycomb\cite{rammal} lattices.
The Hall conductance is quantized when the subbands are filled.

While the topological properties of 2D systems have been intensively studied,
the question of how an applied magnetic field modifies a topological system has
received  less attention\cite{Morais,canadianos}.
Two questions immediately arise:
{\it (i)} How does the chirality introduced by the magnetic field interfere with that from
the underlying band's  topology?
And given that the periodic potential responsible for the topologically nontrivial band
produces a Hofstadter spectrum under a magnetic field, with
a quantized charge Hall conductance when the chemical potential lies in the band gaps,
then {\it (ii)}
How does the non-trivial topology of the underlying lattice
modify this quantization of the Hall conductance?

The effect of a weak magnetic field in a two-band, $C=1$ Chern insulator
has been considered by Haldane in his seminal paper \cite{haldane1988}.
The pair of $n=0$ Landau levels, one per Dirac cone, is degenerate in this
model, in contrast to a trivial system. This fact was used by Haldane to
prove the existence of an anomalous Hall insulator (AHI) in the limit of zero 
magnetic field. 
Although  Haldanes's original construction was devised for  the honeycomb lattice,
we  here consider its general application  to  any time-reversal invariant 
lattice model with arbitrary number of Dirac cones and refer to it as ``Haldane's AHI''. 
It is the effect of a perpendicular magnetic field applied to this generalized
Haldane AHI that we address here.

In Sec.~\ref{sec:model} we review Haldane's construction of the AHI in a way
that can be generalized to models with arbitrary Chern number.
In Sec.~\ref{sec:weak} the effect of a weak magnetic field on the 
AHI is considered
and explicit lattice models are presented. 
The effect of the magnetic field on the 
Kane-Mele $\mathbb{Z}_2$-topological insulator is also addressed.
In Sec.~\ref{sec:strong} 
the Hofstadter spectrum for the AHI in a  magnetic field is presented.
A final summary is given in Sec.~\ref{sumario}.

\section{Generalized Haldane AHI}
\label{sec:model}

 To model a spinless topological Bloch band,  
a two-sublattice system, at least, is needed \cite{SunFradkin2008}.
The Hamiltonian that contains the minimal ingredients can be written as
\begin{equation}
\hat H = \boldsymbol h( \boldsymbol k) \cdot  \boldsymbol \tau
\label{Hk}
\end{equation}
where the Pauli matrices
$\tau_\mu$ ($\mu=1,2,3$) act on the  sublattice 
(``pseudo-spin'') space and $\boldsymbol k=(k_x,k_y)$ runs over the 
Brillouin Zone (BZ).
There are points in the BZ where the gap closes when a topological transition occurs.
Right at the transition, the spectrum at such points is a Dirac cone \cite{haldane2004}, 
and the opening of a gap is due to a finite Dirac mass, $h_z$. 
Suppose that at some point $\boldsymbol K$ in the BZ the Hamiltonian can be linearized as 
\begin{equation}
\hat H \approx \left( -i\hbar v_F\partial_x,
 -i\hbar v_F\partial_y,h_z   \right)\cdot \boldsymbol \tau 
 \label{coneK}
\end{equation}
The contribution of this Dirac point to the Chern invariant of the lower band, 
\begin{eqnarray}
C &=&  \frac{1}{4\pi}\int dk_x\ dk_y \  \frac{\partial \hat{\boldsymbol h}}{\partial k_x} 
\times  \frac{\partial \hat{\boldsymbol h}}{\partial k_y} \cdot \hat{\boldsymbol h}\,,
\label{chern}
\end{eqnarray}
is  given by
 $\Delta C=\frac 1 2  {\rm sgn}\left[ h_z\left(\boldsymbol K\right)   \right]$.
Time-reversal symmetry (TRS) requires both $h_{x(z)}$ to be  even functions
of $\boldsymbol k$ and $h_y$ to be  odd. 
Therefore, another Dirac point must exist at point $\boldsymbol K'$ 
related to that at point $\boldsymbol K$ by TRS.
In the vicinity of $\boldsymbol K'$   the Hamiltonian can be linearized as 
\begin{equation}
\hat H \approx \left( i\hbar v_F\partial_x,
 -i\hbar v_F\partial_y,h_z   \right)\cdot \boldsymbol \tau \,,
 \label{coneKl}
\end{equation}
which gives the  contribution  $-\frac 1 2  {\rm sgn}\left[ h_z\left(\boldsymbol K'\right)   \right]$ 
to the Chern invariant \eqref{chern}.
Since TRS imposes the Dirac masses to be the same,
$h_z\left(\boldsymbol K\right)=h_z\left(\boldsymbol K'\right)$, 
the total Chern number is $C=0$, and a topologically trivial insulator
is realized.

The above pair of Dirac cones embodies the fermion's doubling theorem \cite{NN}.
In order to have nonzero $C$, TRS must be broken. 
Following a procedure analogous to that of Haldane's \cite{haldane1988}, we may opt to break TRS 
by choosing the Dirac masses as $h_z\left(\boldsymbol K\right)  = - h_z\left(\boldsymbol K'\right)$
and the lower band's Chern number is then $C={\rm sgn}[h_z\left(\boldsymbol K\right)]$. 
The resulting system is the Haldane's AHI.
In a finite system, edge states have the chirality shown in 
figure~\ref{chiralities}(a).
The Dirac masses determine the sign of $C$, hence the edge states' chirality.

Model bands with arbitrarily higher $C$  may be constructed, having  $|C|$ such pairs of Dirac cones
in the BZ.

\section{Weak field}
\label{sec:weak}
\subsection{Low energy, continuum description}

\begin{figure}
\begin{centering}
\includegraphics[width=6cm]{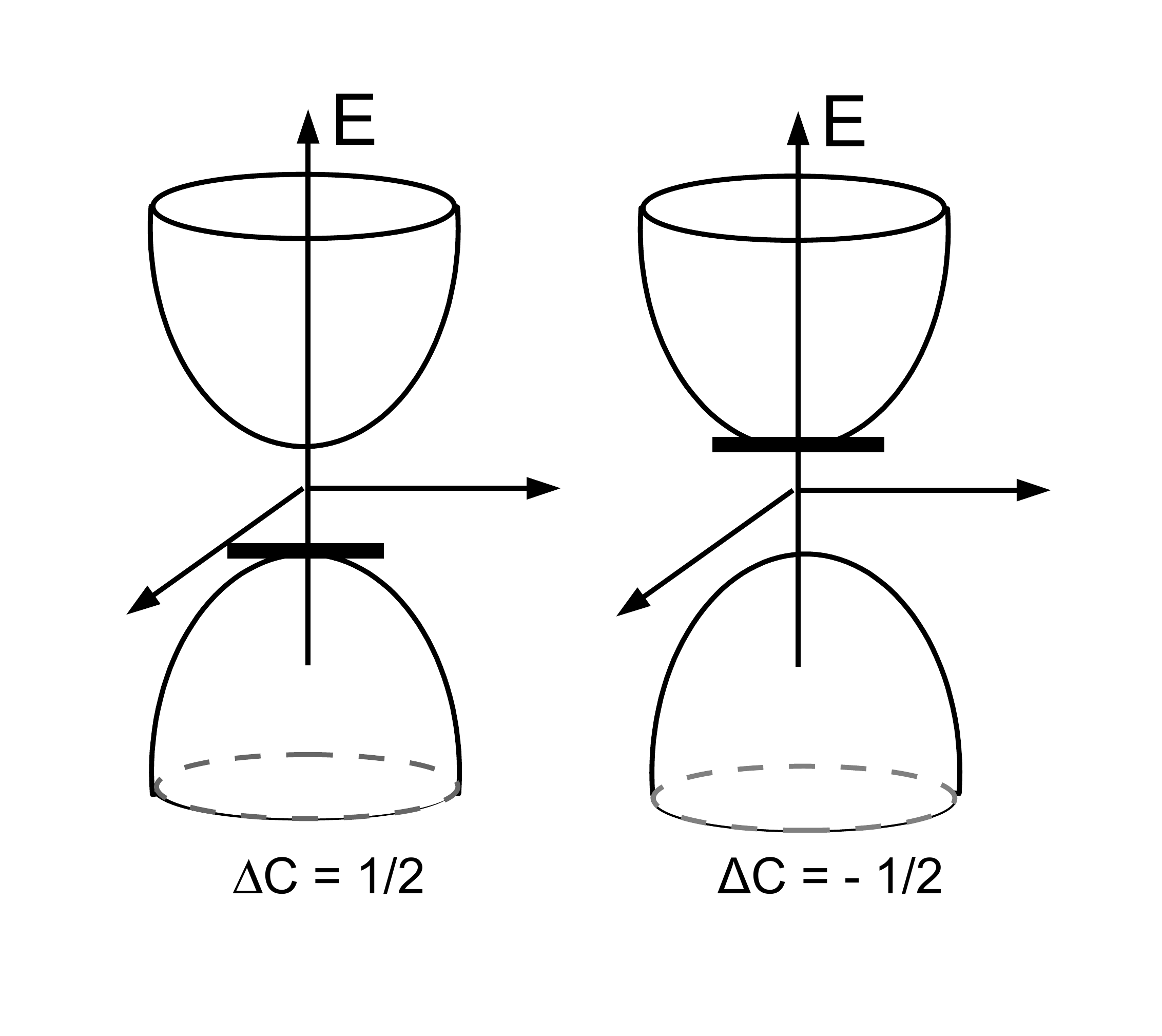}
\includegraphics[width=6cm]{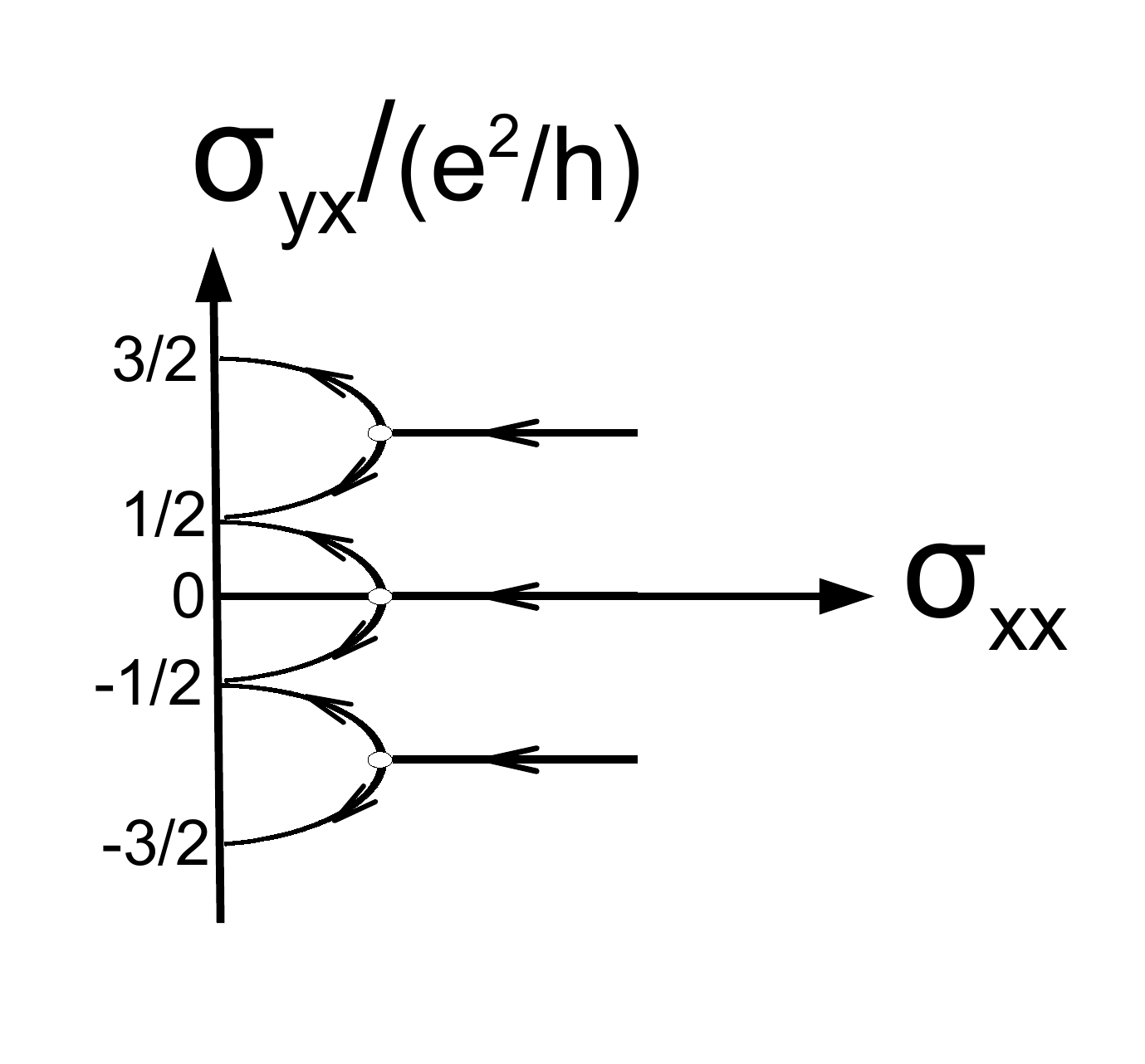}
\par\end{centering}
\caption{\label{conesB}(color online). 
{\bf Top:} The position of the $n=0$ LL (thick horizontal bar)
with respect  to the massive Dirac cones if $B>0$. 
The Chern number of the lower band of a single cone 
(in the absence of magnetic field) is also indicated.
If $B<0$ the LL's positions are interchanged.
{\bf Bottom:}  
Renormalization flow of conductivities as the system size increases. 
This graph proposes a generalization of the
previous scaling theory of Anderson 
localization \cite{khmelnitskii,huckestein} to a single Dirac cone.} 
\end{figure}

We now consider the effect of an applied  perpendicular magnetic field, for spinless particles, 
from the minimal coupling to the vector potential
$-i\hbar\nabla \rightarrow -i\hbar\nabla - e\boldsymbol A$
 in Eqs. \eqref{coneK} and \eqref{coneKl},  
where $\nabla\times\boldsymbol A=B\hat z$. 
The Hamiltonian matrix  now has the form:
\begin{equation}
\hat H = \left(\begin{array}{cc}
h_z & \hat{\cal O}\\
\hat{\cal O}^\dagger & - h_z
\end{array}\right)\,,
 \label{HB}
\end{equation}
where $\left[ \hat{\cal O},  \hat{\cal O}^\dagger \right]=-2\hbar v_F^2eB$ at cone $\boldsymbol K$
and $\left[ \hat{\cal O},  \hat{\cal O}^\dagger \right]=2\hbar v_F^2eB$ at cone $\boldsymbol K'$. 
The spectrum  consists of Landau levels (LLs),
with energies given by  $E_n=sgn(n) \sqrt{h_z^2 + |n| 2\hbar v_F^2\left|eB\right|}$ where the relative integer $n \neq 0$.  
For $n \neq 0$ the LL distribute themselves symmetrically around zero energy. But the position 
of the $n=0$ LL depends on the cone's chirality and the sign of  $B$, according to
\begin{eqnarray}
 \left[ \hat{\cal O},  \hat{\cal O}^\dagger \right] >0\ &\Rightarrow&\ E_0=-h_z\,,\\
 \left[ \hat{\cal O},  \hat{\cal O}^\dagger \right] <0\ &\Rightarrow&\ E_0=h_z\,.
\end{eqnarray}
We still use $C$ to denote the Chern number {\it before}
the field is applied. 
As usual, each LL contains $\left| eB \right|/h$ states per unit area.
The interplay between the magnetic field and topology  manifests itself  in the position of 
 the $n=0$ LL  in a single Dirac cone, which is 
$E_0= - \left| h_z\right| sgn\left( C\cdot B \right)$, 
as shown in  figure~\ref{conesB} (top).
The Hall response of Dirac fermions has been studied in detail
in the context of  graphene\cite{geim,hatsugai}.
It is known that
each filled LL in a single Dirac cone gives a contribution 
$sgn(B) \frac 1 2 e^2/h$
to the charge Hall conductivity, $\sigma_{yx}$. If the $n=0$ LL
is empty, then $\sigma_{yx}=-sgn(B)\frac 1 2 e^2/h$
for a single cone.

For the trivial insulator with $C=0$ the 
pair of Dirac cones at $\boldsymbol K$ and $\boldsymbol K'$,
which are related by TRS,  have opposite chiralities. When $B \neq 0$  their 
$n=0$ LL's shift in opposite directions, as is shown in figure~\ref{conesB} (top). 
 In thermal equilibrium the fermions migrate to the cone with LL
 energy $E_0=-\left| h_z \right|$,
which becomes completely filled. 
The contribution of such filled
cone to the charge Hall conductivity, $\sigma_{yx}$, is  $\frac 1 2 e^2/h$.
The other cone, which has an empty $n=0$ LL, contributes $-\frac 1 2 e^2/h$.
The total $\sigma_{yx}=0$, so the system remains  a trivial  insulator.

In the case of Haldane's AHI ,
 the two cones at  $\boldsymbol K$ and $\boldsymbol K'$
have the same chirality. Their $n=0$ LL's shift in the same direction:
the cones are perfectly degenerate in a magnetic field. The $n=0$ LL
is half-filled in each cone, and the system is a metal.  The Hall conductivity is not quantized. 
Without considering other effects, we then reach the conclusion that in this clean, ideal situation, 
{\it   Haldane's AHI
becomes  metallic under a perpendicular weak magnetic field, at constant particle density.}
Other effects, such as disorder and interactions, can play a decisive role, however. 
If the Dirac masses are not exactly symmetrical,
one of the LL's becomes  full and the other empty, in which case, the combined $\sigma_{yx}=0$.
This result may also be obtained if we consider the role of disorder and  appeal to the scaling theory 
of localization  in the integer quantum Hall effect\cite{khmelnitskii,huckestein,EM08}. 
Generalizing this theory to a single Dirac cone, the Hall conductance scales
to half-integer multiples of $e^2/h$, as shown in figure~\ref{conesB} (bottom). 
In the present case, if  the two $n=0$ LL's have different fillings, 
one expects the Hall conductance of one Dirac  cone to scale  to $\frac 1 2 e^2/h$ 
while  the other scales to $-\frac 1 2 e^2/h$.
Surprisingly, such an  insulating state arises  independently of whether the chirality of 
$B$ is the same or opposite to the
system's original chirality (figure \ref{chiralities}).  Electron repulsion could also play an important
role, since the $n=0$ LL in a single cone occupies only one of the sublattices.

\subsection{Specific lattice models}

To illustrate what has been discussed in terms of the low energy
effective description (pairs of Dirac cones), 
we consider in this work two-band models on the square lattice 
with $C \geq 1$. For 
comparison we also show results for the original Haldane 
model \cite{haldane1988} proposed for the honeycomb lattice 
(see Sec.~\ref{sec:strong}). 

To be specific,
we introduce a $C = 1$ model on the square lattice where the 
vector $\boldsymbol h( \boldsymbol k)$  in Eq.~\eqref{Hk} reads
(in units where the lattice constant $a = 1$)
\begin{eqnarray}
h_x(\boldsymbol k)&=& -1 + 2\cos(k_x) + 2\cos(k_y)\nonumber\\
h_y(\boldsymbol k)&=& 2 \sin(k_x + k_y)\nonumber\\
h_z(\boldsymbol k)&=& - \frac{1}{2} \sin(k_x)\,.
\label{modelC1}
\end{eqnarray}
The $\boldsymbol K$ Dirac cone is  located at $K_x= -K_y=\cos^{-1}(1/4)$
and $\boldsymbol K' = -\boldsymbol K$ \cite{footnote}. 

We provide the vector $\boldsymbol h( \boldsymbol k)$  
in Eq.~\eqref{Hk} for the Haldane model in the honeycomb lattice, 
which we use below for the sake of comparison,
\begin{eqnarray}
h_x(\boldsymbol k)&=& 1 + 2 \cos(k_x/2) \cos(\sqrt{3}k_y/2)\nonumber\\
h_y(\boldsymbol k)&=& 2 \cos(k_x/2) \sin(\sqrt{3}k_y/2)\nonumber\\
h_z(\boldsymbol k)&=& 2 t_2 [2 \sin(k_x/2) \cos(\sqrt{3}k_y/2) - \sin k_x] \,.
\label{modelH}
\end{eqnarray}
As is well known, the Haldane model has $C=1$ for $t_2 \neq 0$.

We point out that  
one may also construct spinless  Chern insulators not following Haldane's procedure of breaking TRS at the
level of $h_z$.   As an alternative to equations (\ref{coneK}) and (\ref{coneKl}),
consider, for instance, that in the vicinity of  momenta $\boldsymbol K$ and  $-\boldsymbol K$
the Halmiltonian takes the linearized form:
 \begin{eqnarray}
{\boldsymbol K}&:&  \hat H \approx \left( -i\hbar v_F\partial_x,
 -i\hbar v_F\partial_y,h_z   \right)\cdot \boldsymbol \tau \nonumber\\
 -{\boldsymbol K}&:&  \hat H \approx \left( i\hbar v_F\partial_x,
 i\hbar v_F\partial_y,h_z   \right)\cdot \boldsymbol \tau 
 \label{hybreak}
\end{eqnarray}
where $h_z(\boldsymbol K)=h_z(-\boldsymbol K)$.
The term $h_y$ breaks TRS in (\ref{hybreak})  and $C=sgn(h_z)$ in the lower band.
In both Dirac cones the $n=0$ LL has energy $E_0=-h_z$ and all of the previous discussion remains valid.

Such a model 
was proposed in Ref.~\onlinecite{beijing}, for the square lattice,
and can be tuned between $C=1$ and $C=2$  by varying hopping parameters. 
The  vector
 $\boldsymbol h( \boldsymbol k)$  in Eq.~\eqref{Hk} reads
\begin{eqnarray}
h_x(\boldsymbol k)&=& \sqrt{2}(\cos k_x + \cos k_y)\nonumber\\
h_y(\boldsymbol k)&=& \sqrt{2}(\cos k_x - \cos k_y)\nonumber\\
h_z(\boldsymbol k)&=& \frac{1}{4} \sin k_x \sin k_y + 
\frac{t_1'}{2}  (\sin k_x + \sin k_y )\,.
\label{modelC2}
\end{eqnarray}
For $t_1' < 1/4$ we have $C=2$, while $t_1' > 1/4$ implies $C=1$. 
There are two pairs of Dirac cones. 
One of such pairs has the cones located  at 
$\boldsymbol K_1 = \pm (\pi / 2, \pi /2)$  
and the other pair  at 
$\boldsymbol K_2 =\pm (\pi / 2, -\pi /2)$.

\begin{figure}
\begin{centering}
\includegraphics[width=0.49\columnwidth]{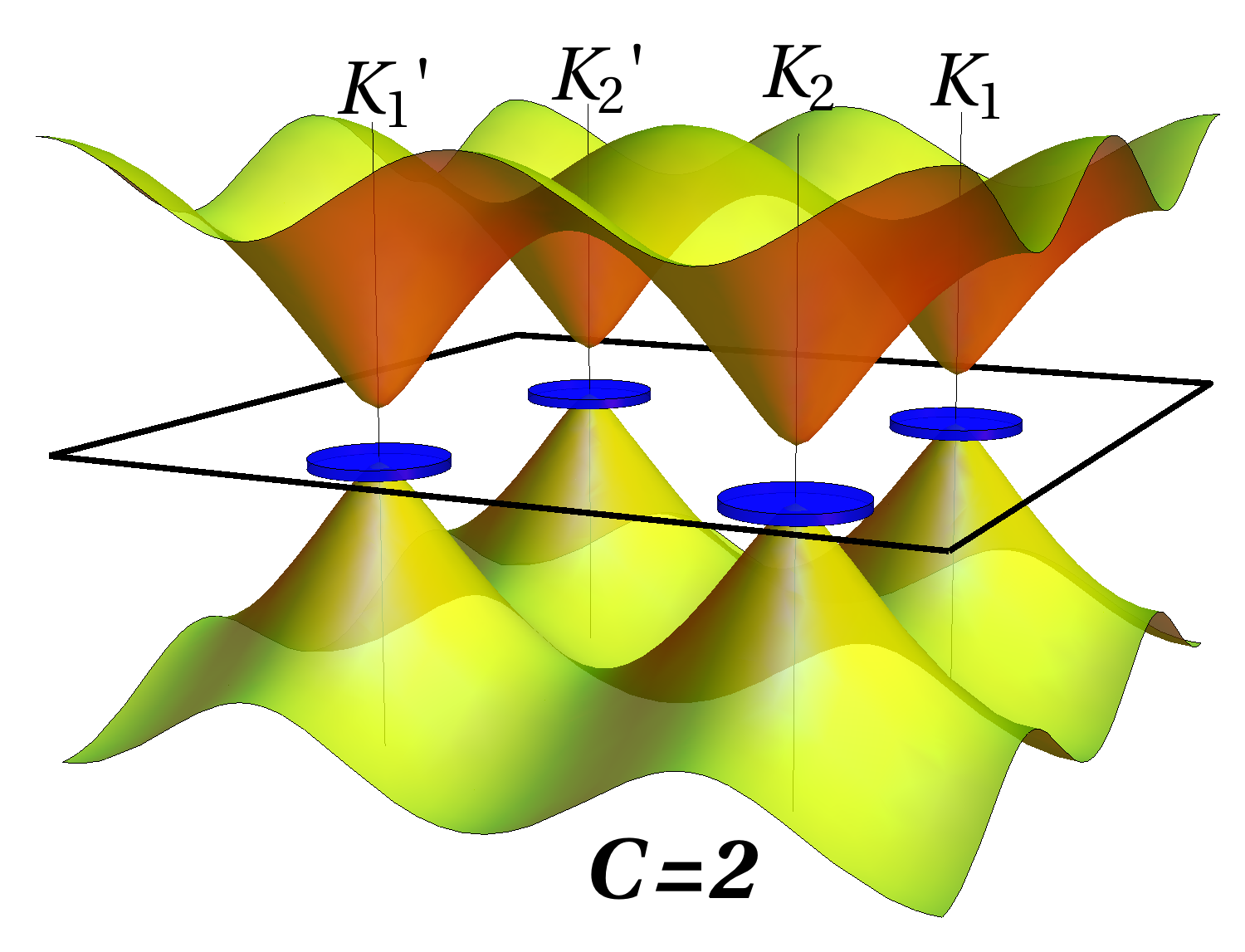}
\includegraphics[width=0.49\columnwidth]{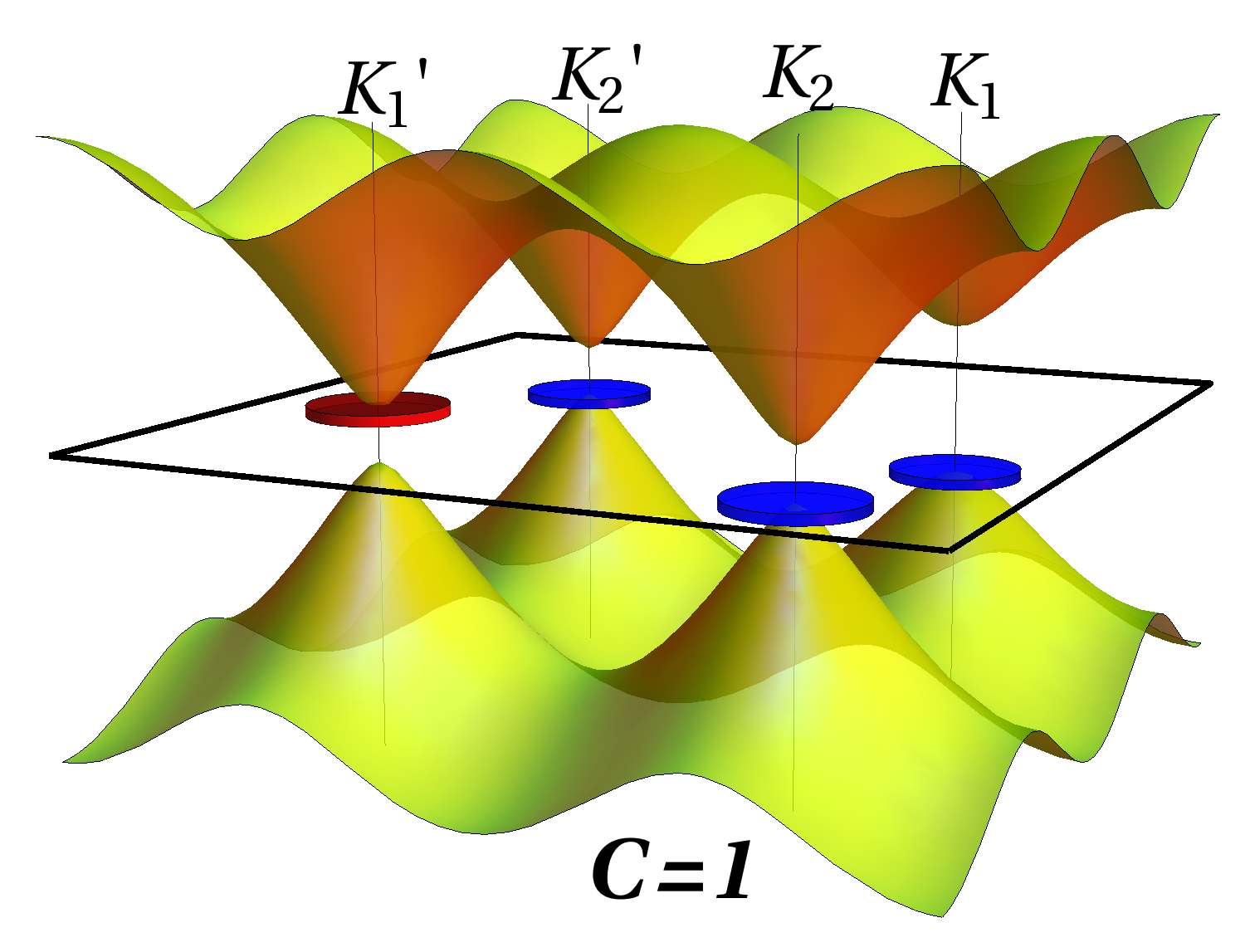}\\
\includegraphics[width=0.95\columnwidth]{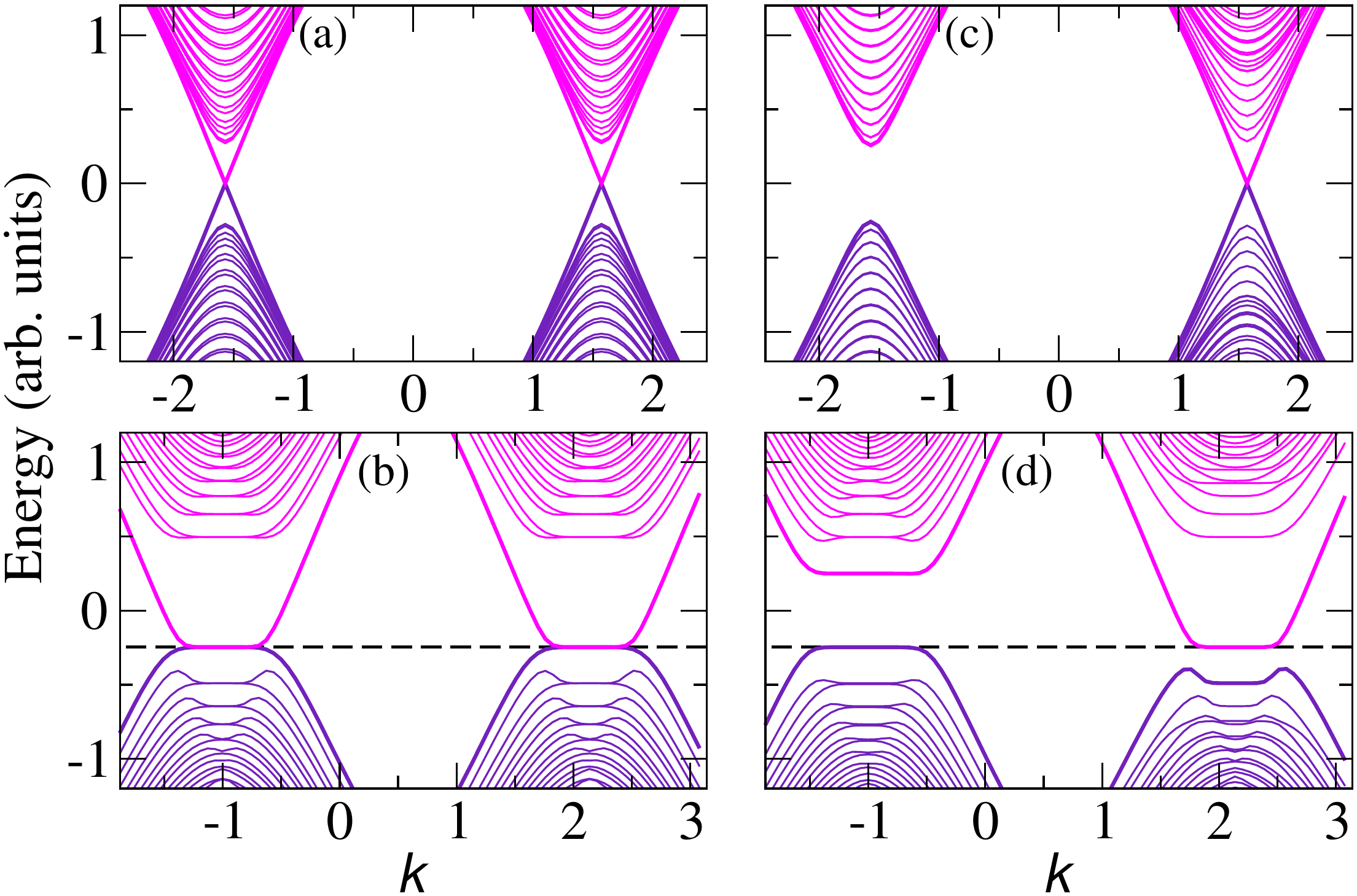}~~
\par\end{centering}
\caption{\label{C2ex}(color online). 
{\bf Top:} The position of the $n=0$ Landau
levels for model~\eqref{modelC2} with $C=2$ (left), and for the same 
model with $C=1$ (right) after band inversion of the $K_1'$ cone. 
{\bf Bottom:}  
Spectrum of model~\eqref{modelC2} in the ribbon geometry for $C=2$ [(a) and (b)]
and $C=1$ [(c) and (d)]. For (a) and (c) the magnetic field is zero, 
for (b) and (d) a weak magnetic field with a flux per square 
lattice unit cell of $\phi/\phi_0 \approx 3.7\times 10^{-3}$ is present.
The $n=0$ LLs are indicated.
The horizontal dashed line marks the Fermi level at half-filling.} 
\end{figure}

\subsubsection*{Case study}

As an example  let us examine the
$C=2$ lattice model~\eqref{modelC2}. 
 The band structure for $B=0$ is
shown in figure~\ref{C2ex} (top), both for $C=2$ (left panel) and for
$C=1$ (right panel). 

For finite magnetic field each Dirac cone originates
a single $n=0$ Landau level. For $C=2$ all four $n=0$ Landau levels
are degenerate and, as anticipated above, this Haldane's AHI is matallic
at half-filling. The position of the $n=0$ Landau levels for a particular
choice of the magnetic field sign is shown in the left panel of
figure~\ref{C2ex} (top). 

The system may become a $C=1$ Chern insulator
after band inversion of the $\boldsymbol K_1'$ cone. As shown in the right
panel of figure~\ref{C2ex} (top) the two $n=0$ Landau levels from 
$\boldsymbol K_1$ and $\boldsymbol K_1'$ now move in opposite directions.
At half-filling the $n=0$ Landau level at $\boldsymbol K_1'$ 
has the highest energy and
becomes
empty, while the one at  $\boldsymbol K_1$,
with the lowest energy,
 becomes fully occupied; note that
the gap (Dirac mass) at  $\boldsymbol K_1$ increases when we close and
reverse the gap at  $\boldsymbol K_1'$ by tuning $t_1 '$. 
So, as in the $C=2$ case,
this $C=1$ Haldane's AHI is metallic, and only for $C=0$ would the
system become a trivial insulator. 

A quantitative description can
be obtained in the ribbon geometry shown in figure~\ref{chiralities}(c). 
In figure~\ref{C2ex}(a) and~\ref{C2ex}(c) 
(bottom panel) we show the spectrum at $B=0$ for $C=2$ and $C=1$, respectively.
The number of edge states running at each edge (same velocity) is precisely
$C$, as it should be for a Chern insulator. The $B\neq 0$ case is shown in
figure~\ref{C2ex}(b) and~\ref{C2ex}(d) (bottom panel), respectively
for $C=2$ and $C=1$. 
The magnetic flux per square lattice unit cell, $\phi$, is set to
$\phi/\phi_0 \approx 3.7\times 10^{-3}$, where $\phi_0=h/e$ denotes the flux quantum.  
It is clear that at half-filling
for $C=2$ there are four  $n=0$  Landau levels crossing the Fermi level 
(horizontal dashed line),  while there are only two for $C=1$.
Note also that the Hall conductivity as obtained from the 
Laughlin-Halperin\cite{laughlin,halperin} argument fully agrees with
the contribution expected from Dirac cones in a low energy description.
For $C=2$ the Hall conductivity is $\sigma_{yx} = 2e^2/h$ if
the four $n=0$ Landau levels are full and $\sigma_{yx} = -2e^2/h$ if they
are empty -- the new Fermi level always crosses two edge states per edge.
In the $C=1$ case the Hall conductivity is $\sigma_{yx} = e^2/h$ if
the two degenerate $n=0$ Landau levels are full and $\sigma_{yx} = -e^2/h$ if 
they are empty -- the new Fermi level only crosses one edge state per edge.
This is nothing but $\sigma_{yx}= \pm |C| e^2/h$, as expected.

\subsection{Kane-Mele $\mathbb{Z}_2$-topological insulator}

\subsubsection{Low energy, continuum description}

 It is interesting now to consider Kane-Mele's\cite{kanemele} construction of the topological insulator (TI).
 Endowing the fermions with spin,
the spin $s_z$ particles see the Dirac cones at $\boldsymbol K$ and  $\boldsymbol K'$
with  $n=0$ LL energy $E_0=-|h_z|$ as in figure~\ref{conesB} (top left), while the
 $-s_z$ particles see the two Dirac cones with the energy level $E_0=|h_z|$ shown in figure~\ref{conesB} 
(top right). These  $n=0$ LLs at   $E_0=\pm |h_z|$ are then initially half filled. 
The thermal equilibrium configuration in achieved when 
the electrons migrate to the cone with  lowest $n=0$ LL energy, $E_0=-|h_z|$, which becomes
completely filled with spin $s_z$. Thus the  total system becomes  spin polarized, 
with  total spin density 
$2s_z\left| CeB\right|/h$ since there are $|C|$ pairs of cones. 
Transitions between such spin polarized LL's by optical absorption 
were discussed very recently for the particular case of 
silicene\cite{canadianos} -- the experimental realization \cite{siliceneExp} 
of the Kane-Mele's TI originally proposed for graphene\cite{kanemele}.
The total  $\sigma_{yx}=0$
but there is a finite spin Hall conductance since the $n=0$ LL is filled with $s_z$ electrons 
while  the $-s_z$ electrons fill the $n=-1$ LL of the other cone. 
 Under a magnetic field and at constant electron density,
{\it  Kane-Mele's topological insulator then becomes a 
spin polarized quantum spin Hall insulator}. This conclusion is valid
under the assumption that $s_z$ is conserved, 
so that 
the Chern number matrix\cite{sc} is diagonal
and  $C_{\uparrow} = -C_{\downarrow} =C$.
Such a  state is stable against potential disorder, 
but unstable against  spin-flip perturbations, in which case it 
would  become a trivial insulator;
a similar situation occurs for a quantum spin Hall insulator in a parallel
(in plane)
magnetic field, which breaks both TRS and $s_z$ conservation\cite{nagaosa07}.
Note that the spin polarization is achieved without considering the Zeeman coupling to spin
and stems from the $s_z$-preserving spin-orbit coupling that originated the TI in the first place. When TRS is present,
 the TI's  $\mathbb{Z}_2$ index $\nu$ is given by the parity of $\left| C \right|$. Equivalently, one can use the spin Chern number\cite{sc}, 
$C_{sc}=C_\uparrow -C_\downarrow = 2C$, with
$\nu=(C_{sc}\mathtt{mod}4)/2$.\cite{QWZtheorem,nagaosa07}
 The magnetic field breaks TRS, restoring the  $\mathbb{Z}$ index,  $C$, which counts
the number of edge states for each spin projection running in a given edge. 
Therefore, if  $|C|>1$,  the spin Hall conductance $\sigma_{yx}^s=|C| 2s_z e/h$.   

\subsubsection{Specific lattice model}

\begin{figure}
\begin{centering}
\includegraphics[width=1.1\columnwidth]{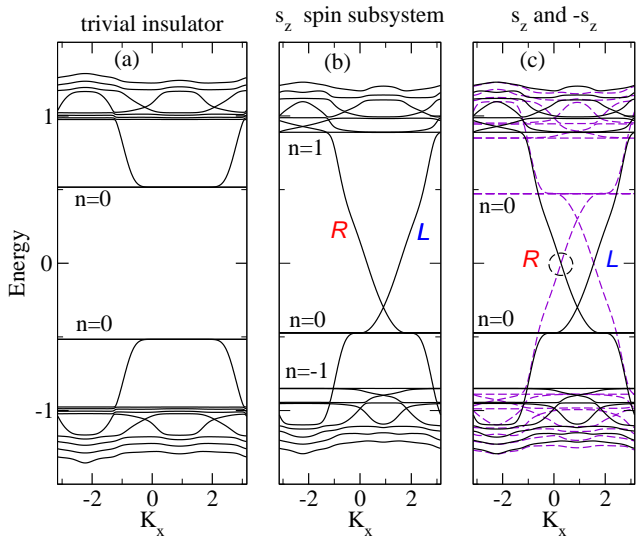}
\par\end{centering}
\caption{\label{edgestates}(color online). 
The spectrum of model \eqref{modelC1} in a  weak magnetic field (flux $\phi/\phi_0 =1/31$
per unit cell) 
and ribbon geometry,  against longitudinal momentum. 
The model has been generalized to include spin 
as explained in the text.
The edge states'  labeling,  $L$ or $R$, follows the convention of figure \ref{chiralities}(c). } 
\end{figure}

Here we consider model~\eqref{modelC1} generalized to include spin using 
the Kane-Mele\cite{kanemele} construction, 
by replacing $h_z$ in (\ref{modelC1}) with 
$h_z(\boldsymbol k)= - sgn\left(s_z\right)\frac 1 2 \sin(k_x)$.
 A weak magnetic field  is  applied perpendicularly. 
figure~\ref{edgestates} shows the LL's and edge states
for the ribbon geometry in figure~\ref{chiralities}(c). It can be seen 
that counter-propagating states with spins $s_z$ and $-s_z$ exist when the chemical potential
lies just  above the $n=0$ LL of the $s_z$ subsystem. 
The Laughlin-Halperin\cite{laughlin,halperin} argument clearly implies the 
spin Hall conductivity $\sigma_{yx}^s=2s_z e/h$, consistent with 
the result of the previous section for $C=1$. It is also clear that
the edge states
are not robust with respect to a spin-flip perturbation, even if such a perturbation is time-reversal
invariant (such as a spin-orbit term).
This is easily seen from the fact that the edge states level crossing 
(marked in figure \ref{edgestates}(c))
occurs at non-zero momentum, a non-time reversal invariant momentum.
Such edge states, in the presence of the magnetic field,   are not Kramer's pairs and can, therefore,
be coupled by a  time-reversal invariant perturbation.
Panel (b) can also be seen as the Chern insulator with $C=+1$, as the $n=0$ LL is full.
 Panel (a) shows the trivial insulator obtained by replacing $h_z$ in (\ref{modelC1}) with $h_z = - 1/2$,
 for comparison.

\subsection{Superlattice potential effect}

Now consider   that the above Haldane's AHI in a perpendicular magnetic field 
$B>0$ is  also subjected to a weak square superlattice potential with rational flux 
$\phi/\phi_0 = p/q$ per unit cell. 
The superlattice potential is diagonal in the
pseudospin index and cannot therefore produce intercone scattering.
The $n=0$ LL of each single cone  splits into $p$ subbands.
The Hall conductance for filled  subbands obeys a Diophantine equation\cite{TKNN}. 
When the chemical potential lies in the $r-$th gap of the split $n=0$ LL, 
the quantized $\sigma_{yx}$
for a single Dirac cone  is given by 
$\sigma_{yx} =  \frac{e^2}{h} \left( -\frac{1}{2} + t_r 
\right)$, where $t_r$ obeys the Diophantine equation\cite{TKNN}:
\begin{eqnarray} 
r=sq + tp\,,
\end{eqnarray}
where $|s|\leq p/2$. 

Consider now
the trivial insulator with a pair of Dirac cones with LL's at energies
$E_0=\pm h_z$, as shown in figure \ref{conesB} (top). 
If only the lower cone is filled, then the Hall conductance is
\begin{eqnarray} 
\sigma_{yx} =  \frac{e^2}{h} \left( t_r -1  \right)\,.
\end{eqnarray}
In the case where the Dirac mass vanishes (as in spinless graphene),  
the $n=0$ LL's
are degenerate at $E_0=0$ and 
$\sigma_{yx} =  \left( 2t_r -1  \right)e^2/h $.

Consider now Haldane's AHI where $C'$ pairs of Dirac cones have degenerate 
$n=0$ LL's at energy $E_0=-|h_z|$ which are partially filled, and
$C''$ pairs of cones have  $E_0=|h_z|$. 
The Hall conductance is
\begin{eqnarray} 
\sigma_{yx} =  \frac{e^2}{h}\left[ C'' \left( 2t_r -1  \right) -C'  \right] \,,
\end{eqnarray}
while the total Chern number 
in the absence of magnetic field and  
superlattice potential is  
$C=C'-C''$.

\section{Strong field}
\label{sec:strong}

\begin{figure}
\begin{centering}
\includegraphics[width=10cm]{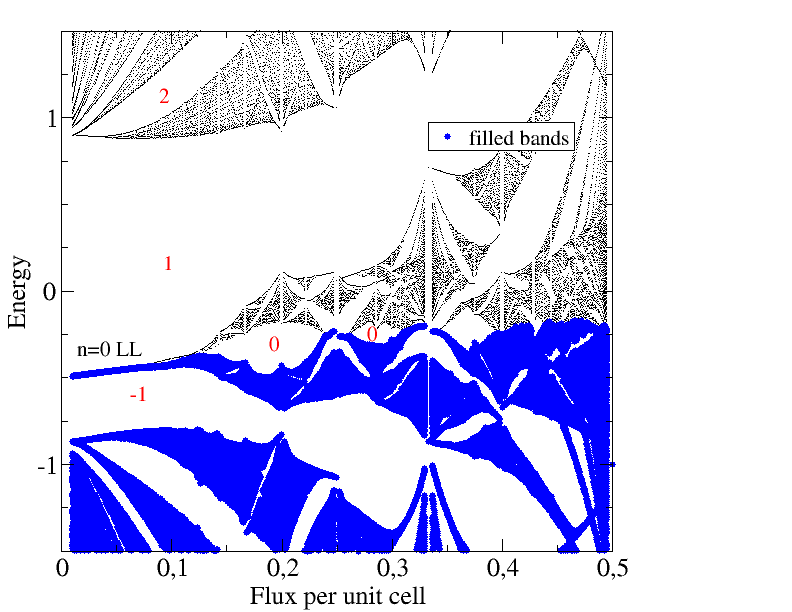}\\
\includegraphics[width=10cm]{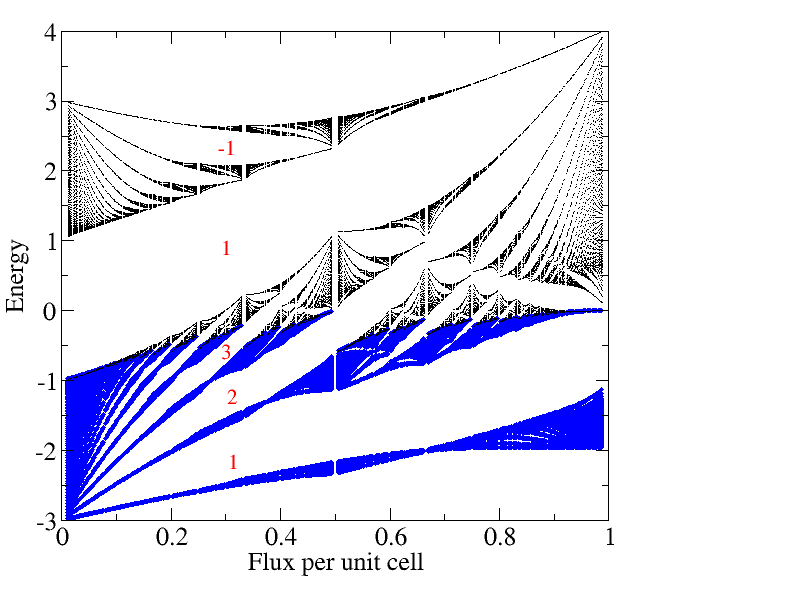}
\includegraphics[width=9cm]{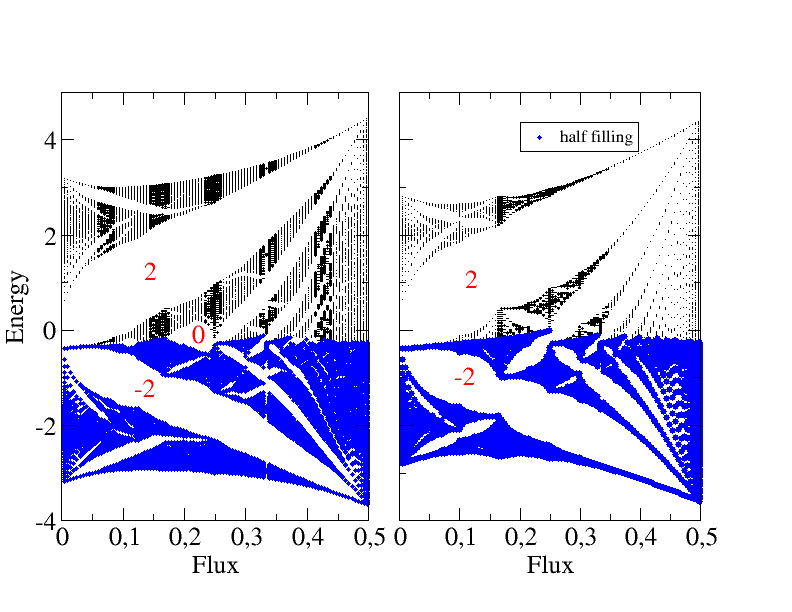}
\par\end{centering}
\caption{\label{borboleta}(color online). 
{\bf Top:} 
The Hofstadter spectrum of model \eqref{modelC1} 
 against flux ($\phi$) per unit cell. 
The occupied bands for a  half filled system are shown, 
as well as the Chern number ($C$) in some of the gaps, 
which gives the the Hall conductance   $\sigma_{yx}$ in units of $e^2/h$.
{\bf Middle}:  The same for model  \eqref{modelH}. 
{\bf Bottom}:  Model (\ref{modelC3}) with $b\neq 0$ (left) and $b= 0$ (right). 
} 
\end{figure}

It is well know that
the spectrum of fermions in a magnetic field and periodic potential 
consists of Hofstadter  bands\cite{hofstadter}. 
In order to study the interplay between  a strong magnetic field's 
gauge potential and a band's topology, 
we use the two-band model in Eq.~\eqref{modelC1} for the square lattice 
which has unit Chern number in the lower band.
The Hofstadter spectrum is displayed in figure~\ref{borboleta}, top panel, 
as a function of the flux, $\phi$, per unit cell.
The spectrum is invariant under the transformation $\phi \rightarrow \phi +1$,
and symmetrical with respect to $\phi=0.5$.
The half-filled band case is shown. The Chern numbers in the gaps
have been calculated with the method given in Ref.~\onlinecite{hatsugai}. 

It is seen in the top panel of figure~\ref{borboleta} that,
 for increasing  flux,   the $n=0$  LL opens and closes a gap 
in the middle,
with zero Hall conductance in the gap.  The system 
goes through  metallic and  trivial band insulator regimes as the
flux per unit cell is increased. 
We note, however, that this feature is model dependent. 
For the original Haldane model proposed in the honeycomb 
lattice,\cite{haldane1988} Eq.~\eqref{modelH}, no such splitting of the  
$n=0$  LL occurs as function of flux. 
This can be seen in the middle  panel of figure~\ref{borboleta}.
The half filled system remains  always metallic, as discussed in the previous 
section. We attribute the different behaviour of the two  models
to the different underlying lattices and the way the two basis atoms 
hybridize  in the lattice. 
While in model~\eqref{modelC1} the two basis atoms
hybridize at the same square lattice site, 
in model~\eqref{modelH} the two atoms are spatially separated. 
As a consequence, there is a Peierls phase for hoppings 
connecting the two basis atoms for the latter case. 
We have verified that when this Peierls factor is artificially suppressed, 
a gap opens at half-filling also for model~\eqref{modelH}.
We may further illustrate this point 
with the following model for a $C=2$ Chern insulator:
\begin{eqnarray}
h_x(\boldsymbol k)&=& 2\sin(k_x) - b \nonumber\\
h_y(\boldsymbol k)&=& 2 \sin(k_y)\nonumber\\
h_z(\boldsymbol k)&=& 0.2 \cos(k_x)\cos(k_y)
\label{modelC3}
\end{eqnarray}
The parameter $b\neq 0$ couples two orbitals at the same lattice site 
and the $n=0$ LL  is split at moderate flux,
as  figure \ref{borboleta} (bottom left) shows.
If $b=0$ the model couples only spatially separated  orbitals. 
Then the  magnetic field
does not split the $n=0$ LL, as  figure \ref{borboleta} (bottom right) shows.
It is therefore expected that Haldane's AHI
in a non-Bravais lattice  under a magnetic field remais metallic 
for all values of the flux.

A final remark is in order.
It has been assumed above that the direct band gap between the bands
is located  at Dirac points. This may not be the case for some nearly flat 
band models \cite{wen,sarma,mudry}. 
In a nearly flat band, the Haldane mass in the Dirac cone becomes large, 
equal to the nearly uniform (across the BZ) gap between the bands. 
The $\boldsymbol k$
 point at which  the direct band gap, $2|\boldsymbol h(\boldsymbol k)|$, is
minimum may happen not to be a Dirac cone, hence the low energy spectrum is not
of the form discussed  [equations (\ref{coneK}) and (\ref{coneKl})] 
and the corresponding analysis of
the n=0 LL no longer applies. Then the lowest LLs may behave as topologically 
trivial LLs do, in which case the AHI placed under a magnetic field
may turn out to remain a band insulator.

\section{Summary}
\label{sumario}
 
In summary, we have studied the effect of a magnetic's gauge field on a topological fermionic band insulator
by considering a generalization of Haldane's model of the  anomalous Hall insulator 
to models with an arbitrary  Chern number.  
We have shown that a spinless system becomes metallic under
a weak  magnetic field, unless other physical effects, such as disorder, are taken into account.
However,  in some model systems a stronger magnetic field can induce an insulating phase.
We have also addressed the effect of a weak square superlattice potential
on the LL splitting and Hall conductance of high Chern number systems.
 In the case of the  $\mathbb{Z}_2$ quantum spin Hall insulator with $s_z$ conservation, 
the magnetic field's vector potential induces a finite magnetization
 even without considering the Zeeman coupling. 
This magnetized quantum Hall state is unstable with respect to 
spin-flip perturbations (including spin-orbit terms).

\section*{Acknowledgments}

We acknowledge the hospitality of CSRC, Beijing, China,
where the final stage of this work has been carried out.
M.A.N.A. would like to thank the hospitality of the University of Gothenburg, Sweden,
where this work was started, and
acknowledges support from the Swedish Foundation for International Cooperation in Research and Higher Education (Grant. No.  IG2011-2028).


\begin{thebibliography}{99}
\bibitem{footnote} 
The spectrum is given by 
$\pm |\boldsymbol h(\boldsymbol k)|$ and the direct band gaps occur
at $\boldsymbol K$ and  $\boldsymbol K'$.
\end{thebibliography}
\end{document}